\documentclass[letterpaper, 10 pt, conference]{IEEEtran}
\IEEEoverridecommandlockouts
\usepackage{cite}
\usepackage{multirow}
\usepackage{amsmath,amssymb,amsfonts}
\usepackage{algorithmic}
\usepackage{graphicx}
\usepackage{textcomp}
\usepackage{xcolor}
\usepackage{siunitx}
\usepackage{subcaption}
\usepackage{comment}
\def\BibTeX{{\rm B\kern-.05em{\sc i\kern-.025em b}\kern-.08em
    T\kern-.1667em\lower.7ex\hbox{E}\kern-.125emX}}

\makeatletter 
\newcommand{\linebreakand}{%
  \end{@IEEEauthorhalign}
  \hfill\mbox{}\par
  \mbox{}\hfill\begin{@IEEEauthorhalign}
}
\makeatother 

\usepackage{eso-pic}
\usepackage{url}
\AddToShipoutPictureBG*{
  \AtPageUpperLeft{%
    \put(0,-40){\raisebox{15pt}{\makebox[\paperwidth]{\begin{minipage}{21cm}\centering
      \textcolor{gray}{This article has been accepted for publication in the proceedings of the 2023 IEEE International Symposium on Medical Measurements and Applications. Content may change prior to final publication. DOI: 10.1109/MeMeA57477.2023.10171940} 
    \end{minipage}}}}%
  }
  \AtPageLowerLeft{%
    \raisebox{25pt}{\makebox[\paperwidth]{\begin{minipage}{21cm}\centering
      \textcolor{gray}{ \copyright 2023 IEEE.  Personal use of this material is permitted.  Permission from IEEE must be obtained for all other uses, in any current or future media, including reprinting/republishing this material for advertising or promotional purposes, creating new collective works, for resale or redistribution to servers or lists, or reuse of any copyrighted component of this work in other works.
      }
    \end{minipage}}}%
  }
}

\begin{document}

\title{Investigation of mmWave Radar Technology For Non-contact Vital Sign Monitoring 
}
\author{\IEEEauthorblockN{
Steven Marty\IEEEauthorrefmark{1},
Federico Pantanella\IEEEauthorrefmark{1},
Andrea Ronco\IEEEauthorrefmark{1},
}

\IEEEauthorblockN{
Kanika Dheman\IEEEauthorrefmark{1},
Michele Magno\IEEEauthorrefmark{1}}\\

\IEEEauthorblockA{\IEEEauthorrefmark{1}Dept. of Information Technology and Electrical Engineering, ETH Z\"{u}rich, Switzerland}
\IEEEauthorblockA{\{martyste, fpantanella, aronco, dhemank, magnom\}@ethz.ch}
}



\sisetup{detect-all}
\maketitle

\begin{abstract}
Non-contact vital sign monitoring has many advantages over conventional methods in being comfortable, unobtrusive and without any risk of spreading infection. The use of millimeter-wave (mmWave) radars is one of the most promising approaches that enable contact-less monitoring of vital signs.
Novel low-power implementations of this technology promise to enable vital sign sensing in embedded, battery-operated devices.
The nature of these new low-power sensors exacerbates the challenges of accurate and robust vital sign monitoring and especially the problem of heart-rate tracking.
This work focuses on the investigation and characterization of three Frequency Modulated Continuous Wave (FMCW) low-power radars with different carrier frequencies of \SI{24}{\giga\hertz}, \SI{60}{\giga\hertz} and \SI{120}{\giga\hertz}.
The evaluation platforms were first tested on phantom models that emulated human bodies to accurately evaluate the baseline noise, error in range estimation, and error in displacement estimation.
Additionally, the systems were also used to collect data from three human subjects to gauge the feasibility of identifying heartbeat peaks and breathing peaks with simple and lightweight algorithms that could potentially run in low-power embedded processors.
The investigation revealed that the \SI{24}{\giga\hertz} radar has the highest baseline noise level, \SI{0.04}{\mm} at \SI{0}{\degree} angle of incidence, and an error in range estimation of 3.45 ± 1.88 cm at a distance of \SI{60}{\cm}.
At the same distance, the \SI{60}{\giga\hertz} and the \SI{120}{\giga\hertz} radar system shows the least noise level, \SI{0.01}{\mm} at \SI{0}{\degree} angle of incidence, and error in range estimation 0.64 ± 0.01 cm and 0.04 ± 0.0 cm respectively. 
Additionally, tests on humans showed that all three radar systems were able to identify heart and breathing activity but the \SI{120}{\giga\hertz} radar system outperformed the other two.
\end{abstract}

\begin{IEEEkeywords}
contactless vital sign monitoring, FMCW, mmWave, radar systems, biomedical systems, comparison
\end{IEEEkeywords}

\sisetup{detect-none}

\section{Introduction}
Monitoring of vital signs, especially  heart rate (HR) and respiratory rate (RR), is essential in modern clinical care~\cite{brekke_value_2019}.
Continuous monitoring of HR and RR can hugely benefit the management of patients by early detection of health deterioration.
Commonly used systems for the measurement of cardiovascular parameters, such as the electrocardiogram (ECG) or the photoplethysmograph (PPG), require contact with the body~\cite{basaklar_wearable_2021}.
Attaching multiple electrodes or placing optical transducers makes current monitoring methods for HR and RR uncomfortable and obtrusive, and is not suitable in many application scenarios.
For instance, contact monitoring is not a viable option in certain patient population groups, such as the pediatric, the geriatric, and those with wounds or burns.
This is due to the added risk of iatrogenic injury or  infection via cross-contamination.
Hence, methods for non-contact monitoring of HR and RR are actively pursued~\cite{singh_multi-resident_2021}.
Non-contact optical solutions have been developed using RGB cameras \cite{selvaraju_continuous_2022}, but these are affected by ambient conditions, such as lighting, temperature and skin colour.
\begin{figure}
    \centering
    \includegraphics[width=0.7\linewidth]{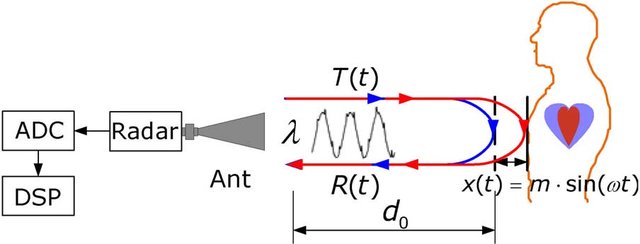}
    \caption{Vital-sign measurement principle of FMCW radars, figure from \cite{stolen_figure}.}
    \label{fig:radar_high_level}
\end{figure}
Millimeter-wave radar technology enables promising and emerging approaches for non-contact vital signs monitoring, circumventing the most pertinent challenges of traditional systems.
Due to the non-contact nature of measurement, there is no risk of infection or injury.
Previous work with radars has shown the feasibility of using such systems for contact-less vital sign monitoring~\cite{singh_multi-resident_2021}.
Depending on the sent-out signal, different radar technologies are defined, such as Continuous Wave (CW), Frequency Modulated Continuous Wave (FMCW), Pulse Coherent Radar (PCR), etc.
Today's technology allows the manufacturing of low-power and accurate FMCW radars, which can enable a new class of non-contact energy-efficient devices.
These devices allow measuring distances and displacements with sub-mm precision, sufficient to capture the small displacements caused by respiration and heart activity, as shown in Figure~\ref{fig:radar_high_level}.
On the other hand, many challenges are still open, and especially HR monitoring with radars is still not fully implemented and evaluated for different subject phenotypes or application scenarios.
Moreover, the previous work is not presenting an investigation and evaluation of the effect of using different carrier frequencies. 

In this paper, we carefully and accurately characterize novel low-power FMCW radars with three different carrier frequencies of \SI{24}{\giga\hertz}, \SI{60}{\giga\hertz} and \SI{120}{\giga\hertz}.
We investigate the choice of carrier frequency and chirp bandwidth to achieve the best performance for vital sign monitoring, given the differences in displacement and range resolution.
The major contributions of this paper are:
1) Characterisation of the baseline noise, range estimation, and displacement error for radars with different carrier frequencies
2) Feasibility of measuring and estimating  HR and RR in three human subjects with different FMCW radars. 
3) Evaluation and discussion of the advantages of the characterized radars and especially the three different carrier frequencies. 

\section{Related Work}
Various studies have sought to determine the most suitable radar system for vital sign monitoring. Depending on the specific vital sign (HR, RR, Tidal Volume or Heart Rate Variability) different radar-based technologies might be appropriate~\cite{islam_radar-based_2022}.
Munoz-Ferreras et al.\cite{munoz-ferreras_doppler_2018} and Giordano et al.~\cite{giordano_survey_2022} emphasize the advantages of FMCW radars over other radar technologies for RR and HR. However, it is uncertain which frequency range is optimal. Lower carrier frequency leads to shallower skin penetration but it only decreases slightly, from \SI{1}{\milli\meter} to  \SI{0.5}{\milli\meter} between \SI{24}-\SI{100}{\giga\hertz}\cite{wu_safe_2015}.
Most approaches use radars with frequencies below \SI{24}{\giga\hertz}\cite{erdogan_microwave_2019} but recent research has shown that higher frequencies provide improvement in measurement~\cite{singh_multi-resident_2021}. Novel systems with larger bandwidths in higher frequencies as \SI{77}{\giga\hertz}~\cite{dai_enhancement_2022} and \SI{120}{\giga\hertz}~\cite{lv_non-contact_2021} showed promising results.
A study on low power radars between \SI{2}{\giga\hertz}-\SI{16}{\giga\hertz} showed that for a given power, an increase in frequency results in a higher sensitivity to small movements~\cite{obeid_low_2008}. 
This suggests that a comparison of higher frequency ranges ($>\SI{16}{\giga\hertz}$) within the same radar technology is necessary to obtain the most power-efficient non-contact vital sign monitoring platform. 

Estimating the HR and RR from the displacement signal involves different signal processing steps.
Respiration causes a large displacement of the chest and hence, relatively simpler signal processing techniques like Fast Fourier Transform (FFT) are used to identify the respiratory rate~\cite{alizadeh_remote_2019}. However, determining HR is more complex due to smaller chest displacements (one order of magnitude)~\cite{shafiq_surface_2015} and confounding factors of breathing rate harmonics. Previous work has used empirical mode decomposition (EMD)~\cite{antolinos_cardiopulmonary_2020} or Random Body Motion Cancellation algorithms~\cite{dai_enhancement_2022} to tackle these problems. However, prior investigations have not evaluated the raw radar signal measured at different carrier frequencies, which might necessitate computationally intensive post-processing to extract vital sign estimations. 

For an initial evaluation, phantoms that mimic human chest movement to characterize different radars have been used. This has been implemented through the use of a metallic pendulum~\cite{dong_doppler_2020} or a vibrating metal plate~\cite{bakhtiari_compact_2012}. The former is restricted to single-frequency movements, whereas the latter can do multi-frequency movements to simulate both the heart and the breathing displacement together. Hence, in this work, a metal plate is used for the evaluation of the characterized radar systems.

\section{Methodology}
To have a fair comparison of how different carrier frequencies affect the signal quality for vital sign monitoring, we evaluate three FMCW radar modules of the same family from Infineon Technologies: BGT24, BGT60, and BGT120.
All radar systems are designed for low-power applications, with an estimated average power consumption of ~\SI{8}{\milli\watt} (sensor only) in our setting at \SI{100}{chirps\per\second}.
The main difference between the three systems is the frequency of the carrier signal (\SI{24}{\giga\hertz}, \SI{60}{\giga\hertz} and \SI{120}{\giga\hertz}) and the chirp bandwidth $B$, which is reported in Table~\ref{tab:radar_freq}.

\subsection{Setup and Configuration}
\label{sub:setup}
For each radar system, one transmitter antenna and one receiving antenna are used.
In order to have a fair comparison, all radars were configured with a sampling rate $F_c$ of \SI{2}{\mega\hertz} and $n = 128$ samples are acquired for each chirp.
A chirp is generated every \SI{10}{\milli\second}.

The slope of the chirp is defined as $S = B/T_c$, with $T_c = n/F_c$.
The range bin resolution is defined as $R = \frac{c}{2B}$, where $c$ is the speed of light. The max range is the product of the number of samples per chirp and the bin resolution: $n \cdot R$.

\begin{table}[]
\caption{Specification of the characterized radar systems}
\centering
\begin{tabular}{l|rrrr}
    Radar & $F_{start}$& $F_{end}$ & $B$ & $R$ \\
    \hline & \\[-2ex]
    BGT24 & \SI{23}{\giga\hertz} & \SI{25}{\giga\hertz} & \SI{2}{\giga\hertz} & \SI{7.5}{\cm} \\
    BGT60 & \SI{58}{\giga\hertz} & \SI{63}{\giga\hertz} & \SI{5}{\giga\hertz} & \SI{3}{\cm} \\
    BGT120 & \SI{116}{\giga\hertz} & \SI{126}{\giga\hertz} & \SI{10}{\giga\hertz}  & \SI{1.5}{\cm}
\label{tab:radar_freq}
\end{tabular}
\end{table}

To evaluate the system performance, a phantom model has been developed to simulate breathing and heart rate pulses.
The device is composed of a square aluminium frame and a flat, movable steel plate with a surface area of \SI{400}{\cm\squared}, comparable to the size of an adult human's chest.
The plate can be moved with high precision by a servo motor controlled by a microcontroller.

Fixed amplitudes for displacement are chosen that are comparable to the displacement amplitude of the chest during cardiopulmonary activity according to~\cite{shafiq_surface_2015}. 
The usual range of heartbeat oscillation on the chest is in the range of \SIrange[range-phrase =--]{0.01}{0.04}{\mm}. Hence, a displacement of \SI{0.08}{\mm} was chosen as a lower bound for the minimum oscillation caused by a heartbeat.
A displacement of \SI{0.03}{\mm} amplitude is in the usual range of heartbeat vibration and 1.2 mm displacement was set as a reference for breathing oscillations, which are in the range of \SIrange[range-phrase =--]{1}{10}{\mm}.
All these displacements are implemented on the phantom with an oscillation frequency of \SI{0.5}{\hertz} and a step size of \SI{0.4}{\micro\meter}.

\begin{figure}[]
\centerline{\includegraphics[clip=true,trim={0cm 10cm 0cm 4.5cm},width=0.6\linewidth]{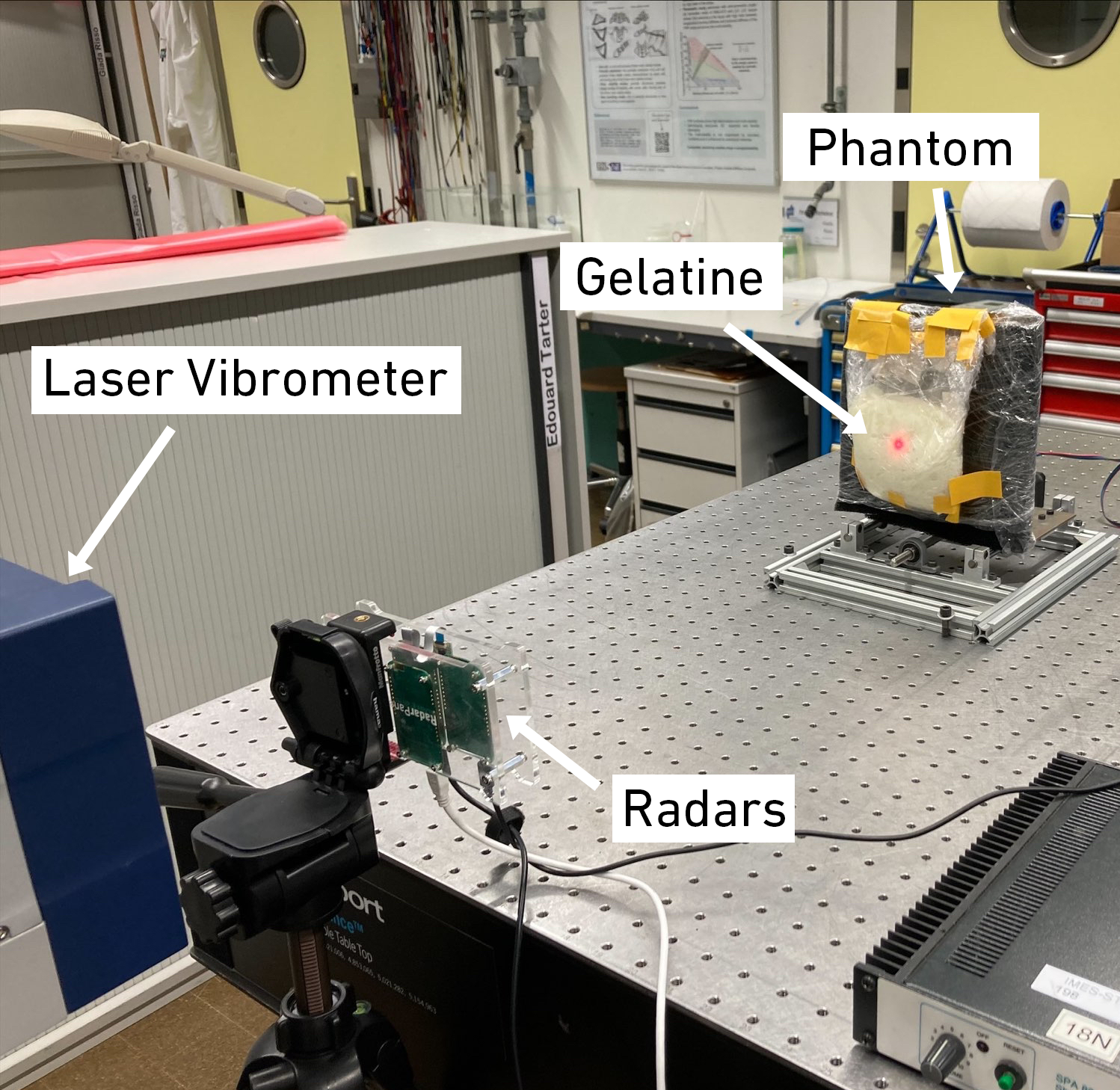}}
\caption{Evaluation setup, featuring the radar systems, the phantom (with gelatin) and the laser vibrometer.}
\label{fig:setup}
\end{figure}

The metal plate has high reflectance in the investigated frequency range of \SI{24}-\SI{120}{\giga\hertz}. Hence, the radars are evaluated in a controlled ideal environment tailored for maximum reflectance from the target object. 
In addition to the bare metal plate phantom, a variant of it was also developed to better emulate the reflectance of the human body.
For this setup, the metal plate was covered with EMI-absorbing foam~\cite{foam}, rated for the range of \SI{5}{\giga\hertz}-\SI{90}{\giga\hertz}, minimizing the reflections from the metal plate.
On top of the foam, a \SI{3}{\cm} layer of gelatinous material prepared accordingly to~\cite{di_meo_tissue-mimicking_2019} was placed. The phantom with gelatin can be seen in Figure~\ref{fig:setup}.

The three radar systems are mounted on acrylic support and held by a tripod, which allows adjusting the position of the radar antennas to be pointed towards the centre of the phantom.
In order to better characterize the displacement, measurements were compared with the Polytec Scanning Vibrometer 500 (PSV500)\cite{vibrometer}, which provided the ground truth on displacement.
The PSV500 can capture vibrational velocities of \SI{0.01}{\micro\meter\per\second} to \SI{30}{\meter\per\second} with integrated conversion to displacement.

\subsection{Signal Processing}
\label{sec:processing}
\begin{center}
    \begin{figure}[]
        \centering
        \includegraphics[width=0.9\linewidth]{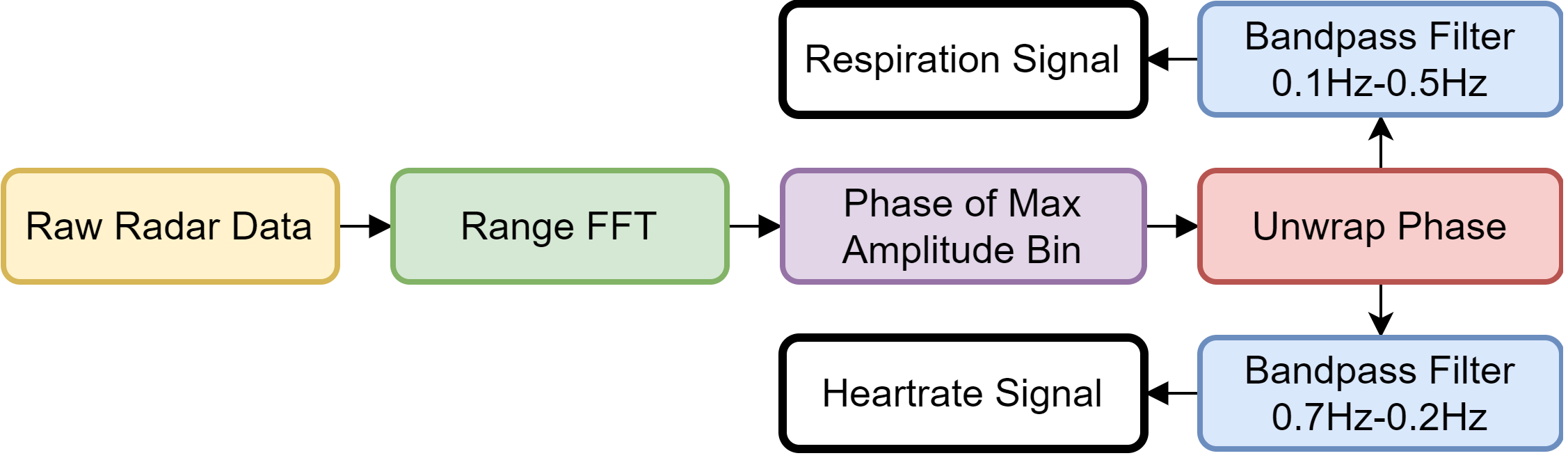}
        \caption{Signal processing pipeline from raw data acquisition to HR and RR estimation}
        \label{fig:processing_pipeline}
    \end{figure}
\end{center}

The task of extracting vital signs from radar data requires complex signal processing.
Since the scope of this paper is a hardware characterization, a previously published pipeline, which is conducive to low-power systems, for HR and RR estimation has been used \cite{paterniani_radar-based_2022} and illustrated in Figure~\ref{fig:processing_pipeline}.

For every chirp, we calculate the spectrum of the raw incoming chirp signal, often named Intermediate Frequency (IF) signal. 
This is done with the FFT, often named Range FFT.
A high-magnitude peak in the Range FFT provides the range (distance) of the target.
The conversion from frequency $f_{IF}$ to distance $d$ is given by:
\begin{equation}
\label{eq:distance}
d= \frac{2S}{c*f_{IF}}
\end{equation}
where $S$ is the slope of the chirp. From the identified target bin, the phase of the signal is extracted, unwrapped and transformed to measure the small displacements of the target. Equation~\ref{eq:displacement} converts the phase change ${\Delta \phi}$ into displacement ${\Delta d}$, where $\lambda$ is the wavelength, indicating the influence of the carrier frequency on the displacement resolution. For higher frequencies, the same displacement is represented by a larger angular change.
\begin{equation} 
\label{eq:displacement}
 \Delta d= \frac{\lambda*\Delta \phi}{4*\pi}
\end{equation}

The displacement signal is band-pass filtered for the heart beat signal between \SI{0.7}{\hertz}-\SI{2}{\hertz} and  for the respiratory signal between \SI{0.1}{\hertz}-\SI{0.5}{\hertz}. These frequency bands correspond to 42-120 bpm and 6-30 breaths per minute, which is a reasonable range for these vital signs~\cite {paterniani_radar-based_2022}. For the purpose of this system characterisation, the heart and respiration rates are obtained from the frequency with the peak amplitude in the FFT of the band-pass filtered signals.

\section{Experiments}
We designed a set of experiments to evaluate the behaviour of the different radar systems for range estimation, displacement estimation, and overall precision in the task of vital sign monitoring.
For the first set of experiments, both phantom models are compared, while vital signs are estimated with data from three human subjects.

\subsection{Range Estimation}
The accuracy of range estimation for all three systems with static targets was evaluated.
The phantom was placed at distances of \SI{30}{\cm}, \SI{40}{\cm}, \SI{50}{\cm}, and \SI{60}{\cm}, and the accuracy and precision with different radars were measured for \SI{60}{\second}.
The distances were chosen to allow for a fair comparison of all three systems while keeping the same settings.
Longer distances would exceed the maximum range for BGT120, making the comparison with this sensor impossible. 
Every recording was divided into 12 sub-intervals of \SI{5}{\second} each, where the distance to the target was estimated  as the average of the detected ranges for the chirps in the sub-intervals.
The ranges are selected by choosing the largest amplitude bin in the Range FFT.
Since the three radar systems offer different bandwidths, the corresponding size of the range bins is also different.
In order to make the comparison fair, zero padding was introduced into the Range FFT calculation to reach a theoretical range precision of \SI{0.157}{\cm} for all three devices.

\subsection{Baseline Noise at Fixed Position}
The variance of the phase for a static target was evaluated, which defines the baseline noise of the radar system.
In this experiment, the phantom was static and placed at a distance of \SI{50}{\cm}. The displacement was extracted from the corresponding bin in the range spectrum. The effect of the orientation of the radars with respect to the target was also evaluated, considering incidence angles of \SI{0}{\degree}, \SI{30}{\degree}, and \SI{60}{\degree}.
For each experiment, \SI{20}{\second} of data were acquired.

\subsection{Displacement Estimation}
To evaluate the accuracy and precision of the displacement estimation, the phantom was set to oscillate with amplitudes of \SI{0.08}{\mm}, \SI{0.3}{\mm}, and \SI{1.2}{\mm} (as described in section~\ref{sub:setup}) with a constant oscillation frequency of \SI{0.5}{\hertz} for \SI{20}{\second}.
The phase from the largest-magnitude bin in the range spectrum is extracted, unwrapped and transformed into displacement.

The ground truth of the displacement is given by the simultaneous measurement of the laser vibrometer, which eliminates slight errors caused by the mechanical imperfections of the phantom model.
Finally, the displacement error is reported as the difference in the peak-to-peak distance measurements between the radar system and the laser vibrometer.

\subsection{Performance on Human Subjects}
Three male human subjects were recorded twice for two minutes using all three radar systems placed 50 cm away. The HR and RR are estimated with the processing pipeline explained in Section~\ref{sec:processing}.
The ground truth was provided by the Polar H10, which is equipped with an ECG sensor and an accelerometer to measure the breathing pattern.
The raw ECG signal of the Polar H10 was processed with the Python library HeartPy\cite{heartpy} to derive the HR ground truth.
The RR was extracted based on the accelerometer data, also acquired by the Polar H10 belt. Feasibility of vital sign monitoring is shown by providing RR and HR estimations along with the Mean Absolute Error (MAE) with respect to the ground truth from the Polar belt.

\section{Experimental Results}
\subsection{Range Estimation and Noise}
\begin{figure}[]
\centerline{\includegraphics[clip=true,trim={0cm 0cm 0cm 3cm},width=\linewidth]{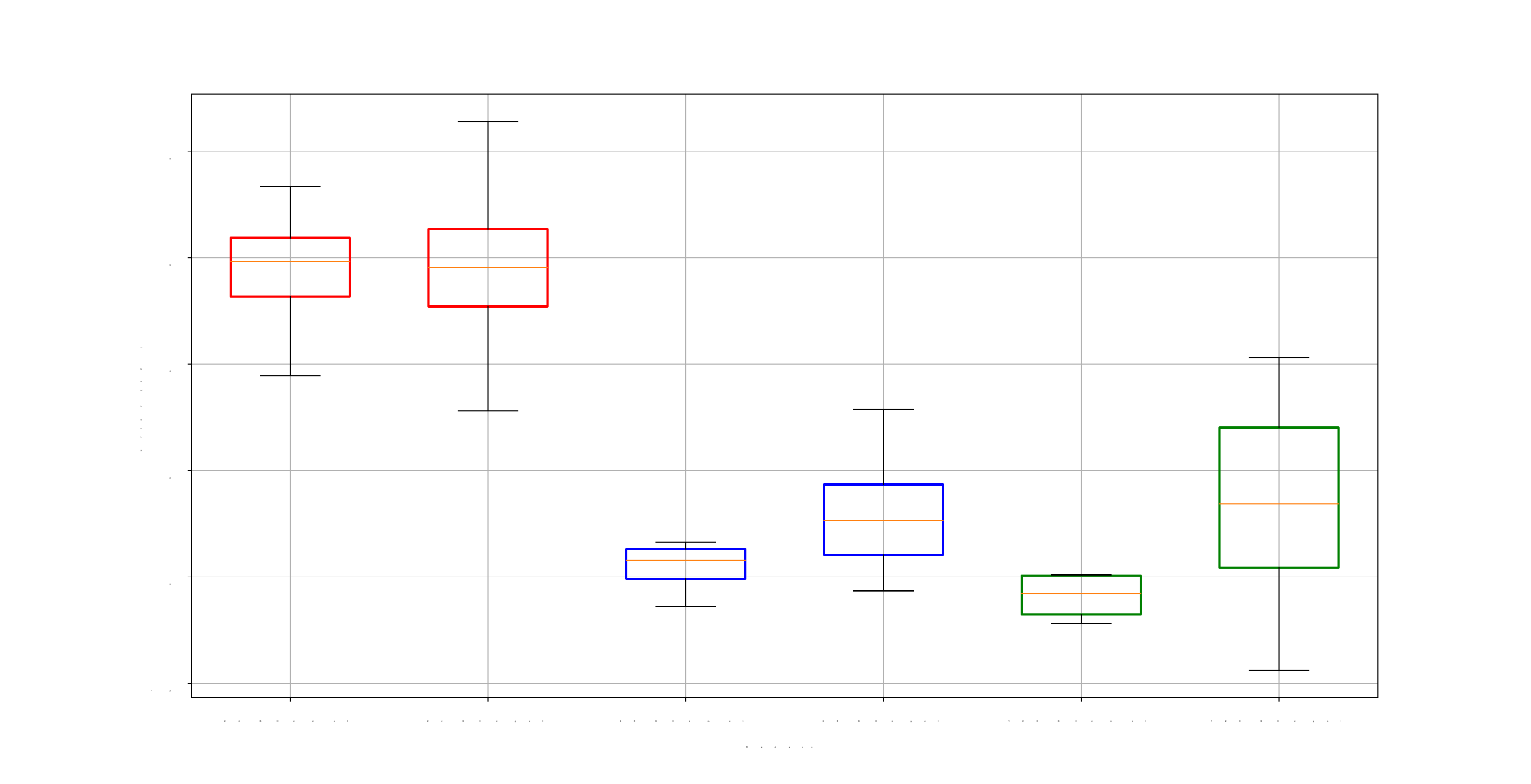}}
\caption{Error in measured range with radars at centre frequencies at 24GHz, 60GHz and 120GHz}
\label{fig:range}
\end{figure}

The errors in estimating the range of the phantom at different distances are shown in Table~\ref{tab:range_error} and summarized over all distances in a boxplot in Figure~\ref{fig:range}.
The BGT24 radar system overestimates the distance by about \SI{6}{\cm} on average for both phantom models.
The BGT60 and BGT120 are more accurate in range estimation with a low (${<1}$ cm) error and variance for the configuration with the metal plate.
For the BGT60 and the BGT120, the gelatine increases the variance of the range measurement and also increases the mean average error of range estimation.
This is expected due to the less ideal reflectance properties of the material.

\begin{table}[]
\centering
\caption{Range estimation error in centimetres.}
\begin{tabular}{l|r|rrr}
Phantom & Range &   BGT24 &    BGT60 &   BGT120 \\
\hline & \\[-2ex]
\multirow{4}{*}{Metal}
 &    \SI{30}{\cm} & $6.25 \pm 0.2$ & $0.49 \pm 0.01$ & $0.02 \pm 0.0$ \\
 &   \SI{40}{\cm} & $5.53 \pm 0.18$ & $-0.55 \pm 0.01$ & $-0.87 \pm 0.01$ \\
 &   \SI{50}{\cm} & $6.57 \pm 0.3$ & $0.14 \pm 0.01$ & $-0.65 \pm 0.01$ \\
 &   \SI{60}{\cm} & $3.45 \pm 1.88$ & $0.64 \pm 0.01$ & $0.04 \pm 0.0$ \\
 \hline & \\[-2ex]
 \multirow{4}{*}{Gelatin}
 &    \SI{30}{\cm} & $6.02 \pm 0.2$ & $1.25 \pm 0.04$ & $0.05 \pm 0.13$ \\
 &   \SI{40}{\cm} & $7.2 \pm 3.2$ & $-0.21 \pm 0.05$ & $-0.01 \pm 0.72$ \\
 &   \SI{50}{\cm} & $6.3 \pm 0.76$ & $0.64 \pm 0.11$ & $2.32 \pm 0.06$ \\
 &   \SI{60}{\cm} & $5.0 \pm 0.18$ & $3.13 \pm 0.01$ & $4.07 \pm 0.04$ 
\label{tab:range_error}
\end{tabular}
\end{table}

\begin{figure*}[t]
    \includegraphics[width=\textwidth]{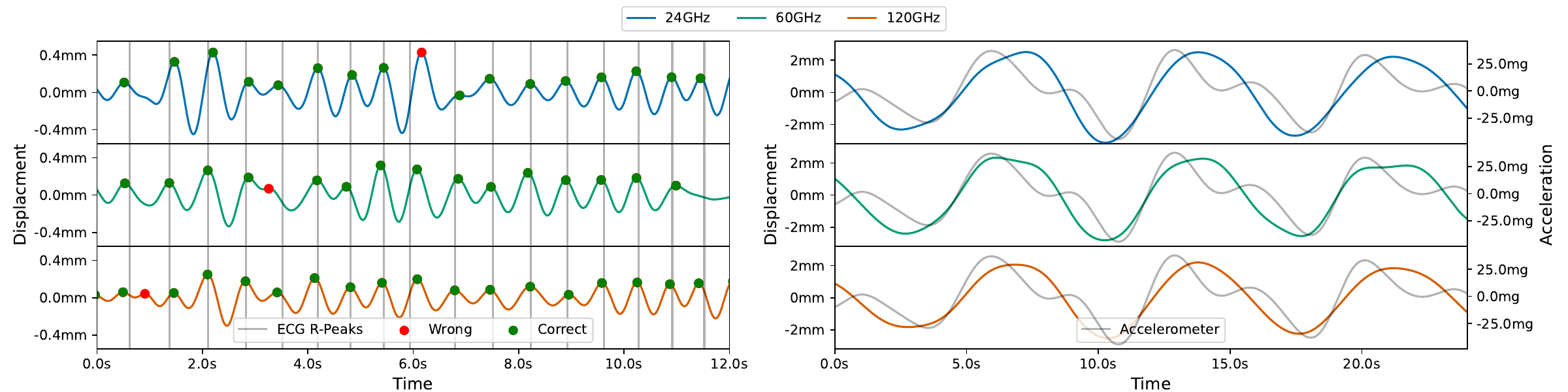}

    \caption{Heart-rate signal with correct and wrongly identified peaks (left) and Respiration signal (right) for all three radar systems.}
    \label{fig:vital_comparison}
\end{figure*}

\subsection{Baseline Noise at Fixed Position}
\label{sec:base_noise}
The baseline noise in each radar system for the two phantom models is given in Table~\ref{tab:basenoise}. The baseline noise recorded from the gelatin phantom is generally slightly higher than that of the metal due to lowered reflectance from the former. Also, the noise levels increase as the angle incidence increases. This can be attributed to ambient sources of noise being captured by the radar from the surroundings. Additionally, the BGT24 has a higher baseline noise than the BGT60 and BGT120 for all angles of incidence and for both phantom models.
\begin{table}[]
\centering
\caption{Baseline noise, variance of the phase in millimeters.}
\begin{tabular}{l|r|rrr}
Phantom & Angle &   BGT24 &    BGT60 &   BGT120 \\
    \hline & \\[-2ex]
     \multirow{3}{*}{Metal}
    &    \SI{0}{\degree} & 0.015 & 0.004 & 0.001 \\
    &   \SI{30}{\degree} & 0.059 & 0.001 & 0.001 \\
    &   \SI{60}{\degree} & 0.044 & 0.021 & 0.348 \\
        \hline & \\[-2ex]

     \multirow{3}{*}{Gelatin}
  &    \SI{0}{\degree} & 0.040 & 0.001 & 0.001 \\
  &   \SI{30}{\degree} & 0.060 & 0.001 & 0.001 \\
  &   \SI{60}{\degree} & 2.988 & 0.031 & 3.796
\label{tab:basenoise}
\end{tabular}
\end{table}

\subsection{Displacement Estimation}

Table~\ref{tab:estdisp} shows the measured displacement of the three radars with respect to ground truth (GT) given by the laser vibrometer. It can be seen that the BGT24 has consistently larger errors in displacement estimation. 
The BGT60 and the BGT120 radar systems show similar errors, however, in the gelatin model, the BGT120 performs better than the BGT60 for the sub-millimetre displacements.
For the BGT60 and BGT120 systems, DC offset removal~\cite{alizadeh_remote_2019} was used right after the range FFT to reduce the influence of clutter from the reflections of static objects.
For the BGT24 system, this technique cannot be used because the small displacements of vital signs (in the range \SI{0.1}{\mm}-\SI{1}{mm}) are too low when compared to the wavelength of the carrier ($\lambda = 12.5$ mm), which makes it difficult to estimate the DC correctly.

\begin{center}
    \begin{figure}[]
        \centering
        \includegraphics[clip=true,trim={0cm 0.5cm 0cm 2.2cm},width=0.9\linewidth]{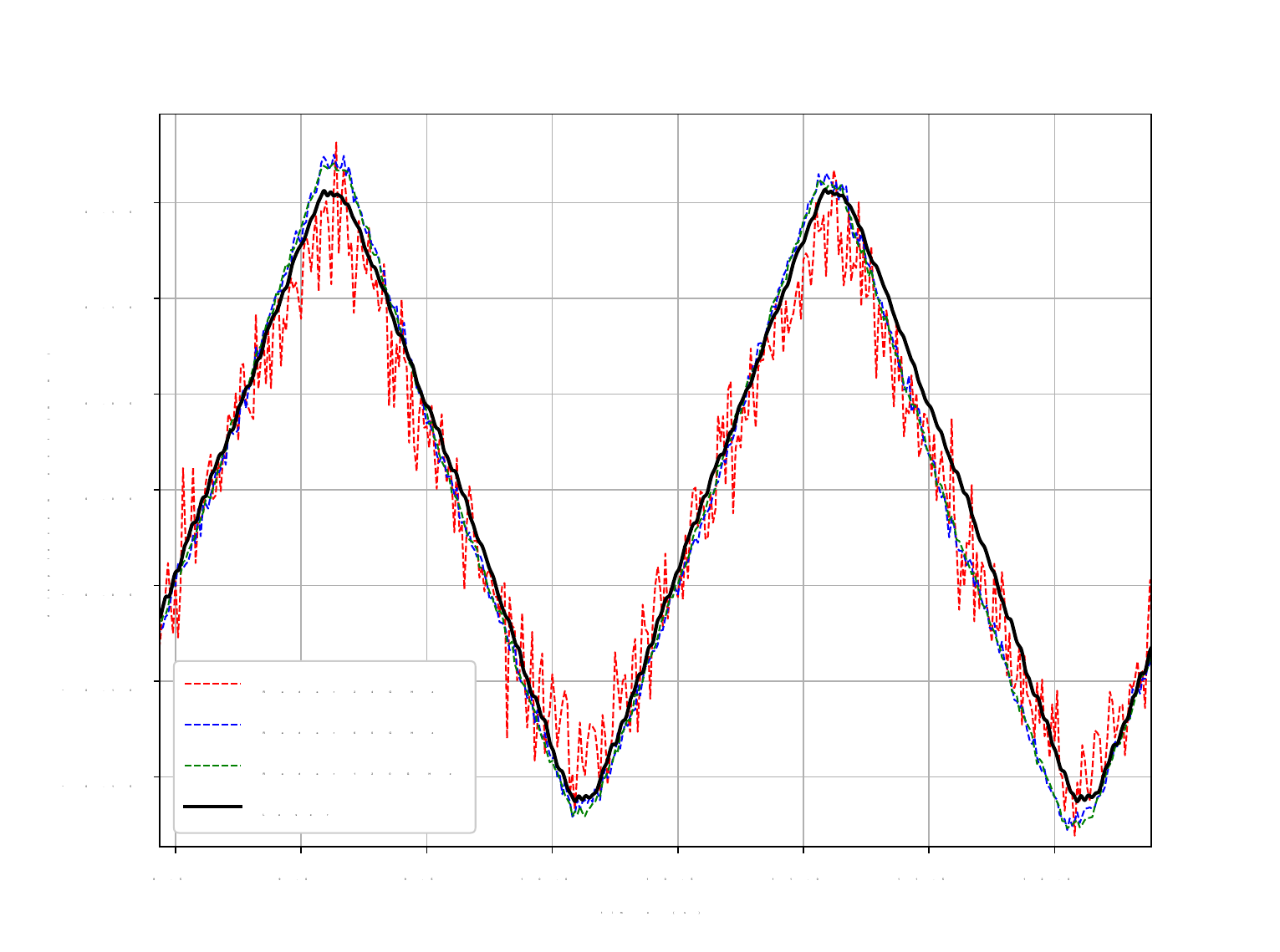}
        \caption{Measured displacement of amplitude \SI{0.3}{\mm}, comparison between the radar systems and the laser vibrometer.}
        \label{fig:displacement_plot}
    \end{figure}
\end{center}

Figure~\ref{fig:displacement_plot} shows that all three systems can track the displacement of the phantom, marked in black.
The BGT24 shows a higher noise level, which is congruent with the results on baseline noise in Section~\ref{sec:base_noise}.

\begin{table}[]
\centering
\caption{Estimated Peak-Peak Displacement absolute error in millimeters}
\begin{tabular}{l|r|r|rrr}
Phantom & Target $\Delta$ & Laser & BGT24 & BGT60 & BGT120 \\
\hline & \\[-2ex]
     \multirow{3}{*}{Metal}
    & 1.2 & 1.187 &  0.038 &  0.018 &   0.028 \\
    & 0.3 & 0.318 &  0.055 &  0.010 &   0.010 \\
    & 0.08 & 0.079 &  0.020 &  0.047 &   0.004\\
    \hline & \\[-2ex]
     \multirow{3}{*}{Gelatin}
  & 1.2 & 1.187 &  0.071 &  0.040 &   0.059 \\
  & 0.3 & 0.313 &  0.120 &  0.033 &   0.013 \\
  & 0.08 & 0.079 &  0.026 &  0.019 &   0.015
\end{tabular}
\label{tab:estdisp}
\end{table}

\subsection{Performance on Human Subjects}
A comparison of the measured displacements from the three radars is visualized in Figure~\ref{fig:vital_comparison}.
On the left, the band-passed signal for the heart rate is shown, where the peaks that appear within a \SI{150}{\ms} window from the actual ECG beat (ground truth) are marked in green.
On the right, the breathing signal is shown, alongside the corresponding Polar acceleration signal.
For this time window, all the radars seem to perform similarly, however, this is not always the case, as we show below.

\begin{figure*}[t]
    \includegraphics[width=\textwidth]{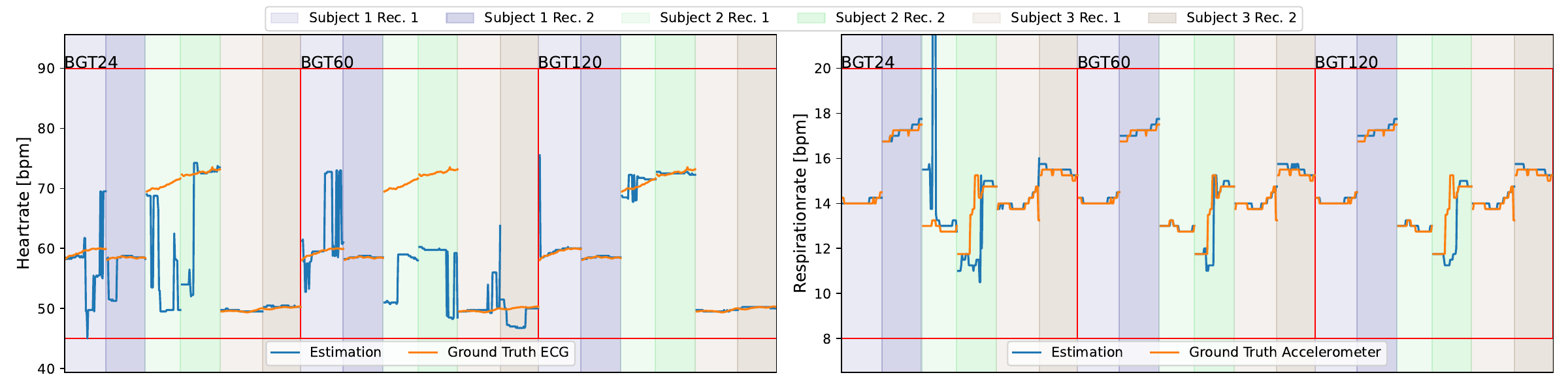}

    \caption{(left) HR estimation and (right) RR estimation from the three different radars for six recordings (two recordings for each subject).  The values are estimated based on a rolling window of \SI{60}{\second} with a step size of \SI{1}{\second}. As the recordings were done simultaneously, the ground truth for each recording (colour) is the same across all radars. }
    \label{fig:vital estimation}
\end{figure*}

For the complete comparison of all recordings, the HR and RR are estimated for windows of \SI{60}{\second}, which are shifted by one second.
The estimations together with the ground truth are visualized in Figure~\ref{fig:vital estimation}. 
Generally, the data suggests that the BGT120 performs the best followed by BGT24 and BGT60.
BGT120 achieves an MAE for the HR of \SI{0.4}{bpm} (beats per minute) with a standard deviation ($\sigma$) of \SI{1}{bpm}, whereas the MAE for the BGT24 and the BGT60 is \SI{4}{bpm} ($\sigma= 7$) and \SI{6}{bpm}  ($\sigma= 7$) respectively.
As visible in Figure~\ref{fig:vital estimation}, all three radars are able to accurately track the HR on some recordings, showing promise for all technologies for this task.
However, for some subjects and recordings, the estimation can be quite off, indicating that this basic algorithm is not sufficiently robust.
Specifically, as mentioned in~\cite{singh_multi-resident_2021}, this algorithm can easily misinterpret the harmonics of the breathing pattern for heart rate.
These issues can be tackled with advanced algorithms, which are not in the scope of this work.

For the RR, all the radars perform similarly with an MAE~${<1}$ breath per minute (brpm). 
This result is expected, as the displacement caused by breathing is around one order of magnitude greater than the one of the heartbeat~\cite{shafiq_surface_2015}.
Only the BGT24 has some noticeable outliers for Subject 2.

\section{Conclusions}
This work characterizes three novel and low-power FMCW radar systems operating at frequencies of \SI{24}{\giga\hertz}, \SI{60}{\giga\hertz} and \SI{120}{\giga\hertz} to evaluate the feasibility of contactless HR and RR monitoring and the benefits of the different frequencies. 
The performance of the radars was evaluated on two phantom systems to estimate the range of the target object, baseline noise, and error in displacement estimation. 
Phantom experiments indicate that the higher frequency (\SI{60}{\giga\hertz} and \SI{120}{\giga\hertz}) systems, which also provide higher bandwidth configurations, show lower error in range and displacement estimation than the \SI{24}{\giga\hertz} radar system. One reason is the frequency-distance (Eq.~\ref{eq:distance}) and the phase-displacement (Eq.~\ref{eq:displacement}) relationship. First, bigger bandwidths allow for finer distance resolution and therefore lower error. Secondly, the same displacement is represented by different IF signal phase changes $\Delta \phi$ depending on the carrier frequency. For higher frequencies, the corresponding phase change is bigger and therefore easier to measure. Thus, a wider bandwidth with higher carrier frequency is beneficial for displacement measurements in the heartbeat range with a high SNR. 
Furthermore, the three systems were also tested on human subjects to investigate the performance in measuring the HR and RR with low complexity signal processing such as FFT.
Tests on 3 subjects showed that the low-power radars could identify the breathing and heart activity patterns. The \SI{120}{\giga\hertz} radar system was most accurate in estimating the HR (\SI{0.4}{bpm}) and RR (~${<1}$brpm). Hence a low-power (\SI{8}{\milli\watt}), radar system was able to achieve high accuracy with a low computational load algorithm for HR and RR estimation. The results are comparable to other state-of-the-art non-contact instruments as cameras (2-8bpm RMSE for HR and 1-4brpm RMSE for RR) without the drawbacks of being affected by lighting or skin colour~\cite{selvaraju_continuous_2022}. 

In the future, research needs to focus on evaluating these low-power radars on a larger sample set to develop algorithms for accurate and precise HR and RR measurement that account for inter-subject variability, due to muscle, adiposity, breast tissue, or body hair on the chest. Additionally, investigating greater distances between the radar and the subject could help in optimising the radar systems for different applications.

\bibliographystyle{IEEEtran}
\bibliography{refs}
\end{document}